\documentclass[aps,pra,twocolumn,superscriptaddress,showpacs,floatfix,longbibliography,citeautoscript]{revtex4-1}

\usepackage[dvips]{graphicx}
\usepackage{color}
\usepackage{latexsym}
\usepackage{amsmath}
\usepackage{amsthm}
\usepackage{amsfonts}
\usepackage{amssymb}
\usepackage{mathptmx}
\usepackage{subfigure}

\usepackage{mathrsfs}

\usepackage{hyperref}
\hypersetup{colorlinks=true}
\usepackage[all]{hypcap} 

\begin{document}

\title{
Detecting Photon-Photon Interactions in a Superconducting Circuit
}

\author{Li-Jing Jin} 
\email{li-jing.jin@lpmmc.cnrs.fr}
\affiliation{Univ. Grenoble Alpes, LPMMC, F-38000 Grenoble, France;\\
CNRS, LPMMC, F-38000 Grenoble, France}

\author{Manuel Houzet}
\affiliation{Univ.~Grenoble Alpes, INAC-SPSMS, F-38000 Grenoble, France;\\CEA, INAC-SPSMS, F-38000 Grenoble, France}

\author{Julia S.\ Meyer}
\affiliation{Univ.~Grenoble Alpes, INAC-SPSMS, F-38000 Grenoble, France;\\CEA, INAC-SPSMS, F-38000 Grenoble, France}

\author{Harold U.\ Baranger}
\affiliation{Department of Physics, Duke University, Durham, North Carolina 27708-0305, USA}

\author{Frank W.\ J.\ Hekking}
\affiliation{Univ. Grenoble Alpes, LPMMC, F-38000 Grenoble, France;\\
CNRS, LPMMC, F-38000 Grenoble, France}

\begin{abstract}
A local interaction between photons can be engineered by coupling a nonlinear system to a transmission line. 
The required high impedance transmission line can be conveniently formed from a chain of Josephson junctions. The nonlinearity is generated by side-coupling this chain to a Cooper pair box. 
We propose to probe the resulting photon-photon interactions via their effect 
on the 
current-voltage characteristic
of a voltage-biased Josephson junction connected to the transmission line. 
 Considering the Cooper pair box to be in the weakly anharmonic regime, we find that the dc current through the probe junction
yields features around the voltages $2eV=n\hbar\omega_s$, where $\omega_s$ is the plasma frequency of the superconducting circuit. The features at $n\ge 2$ are a direct signature of the photon-photon interaction in the system.
\end{abstract}


\date{\today}

\maketitle

\section{Introduction}

Creating strong light-matter interaction attracts increasing attention due to both fundamental reasons \cite{SchoelkopfNat08,HarocheNobel13,WinelandNobel13,QFluidsLightRMP13} and its potential application in quantum communication science \cite{SchoelkopfNat08,NorthupQInfNP14,GisinQCommNP07}. A prototypical system for studying that interaction consists of a quantum system inside a photonic cavity \cite{HarocheRaimondBook}. However, recent rapid experimental advances in several areas \cite{HoiNJP13,vanLooScience13,VersteeghNatCommun14, ArcariPRL14,VetschPRL10,FaezPRL14} have focused attention on \emph{one-dimensional} systems in which the quantum system is embedded in a waveguide or transmission line. In the absence of coupling, photons propagate freely down the line. A coupling between the quantum system and the line generates an effective photon-photon interaction that causes correlations among the photons. This has lead to, for instance, the prediction of Kondo physics \cite{LeClairPRA97}, anti-bunching resulting from a photon-blockade effect \cite{ShenPRL07,ZhengPRA10}, inelastic photon scattering \cite{ShenPRL07,LeHurPRB12,GoldsteinPRL13}, giant Kerr nonlinearities \cite{RebicPRL09}, and entanglement among photons of different frequencies in the line \cite{BeraPRB14-1}. 

The strength of the coupling between the local quantum system and the transmission line has been studied theoretically in detail in the ohmic spin-boson model, which consists of a single two-level system (the spin) bilinearly coupled to the photons in the line (the bosons). It was shown \cite{LeggettRMP87,WeissBook} that the coupling parameter is set by the ratio of the line impedance, $Z$, to the quantum of resistance, $R_Q=h/(2e)^2\approx 6.45\,\rm{k\Omega}$. The impedance of typical transmission lines is of order the vacuum impedance,  $Z_{\rm vac}\approx 377\,\Omega$, 
thereby allowing only weak coupling. 

Superconducting circuits are a promising platform for exploring strong coupling phenomena, and, indeed, the first experiments observing such phenomena have appeared \cite{NiemczykNatPhys10,MooijPRL10}. One benefit of using superconducting circuits is that a chain of Josephson-coupled superconducting islands acts as a transmission line with a large tunable impedance $Z \lesssim R_Q$,  which is only limited by the superconductor/insulator transition \cite{BradleyDoniachPRB84}. Recent experiments have studied the microwave properties of such Josephson junction chains \cite{MaslukDevoretPRL12,BellGershensonPRL12,AltimirasPortierAPL13,WeisslGuichardPRB15}. Moreover, superconducting circuits allow the realization of a variety of quantum systems that behave like artificial atoms \cite{SchoelkopfNat08,SchnirmanRMP01}. 

In our work, we take the quantum system coupled to the transmission line to be a Cooper pair box \cite{BouchiatDevoret98,NakamuraNat99}. 
  Then, we propose to detect the photon-photon interaction generated by that system by measuring the dc current-voltage characteristic of an additional Josephson junction connected to the transmission line. According to dynamical Coulomb blockade theory [also called $P(E)$-theory] \cite{IngoldNazarov92}, Cooper pairs can tunnel incoherently through that probe junction provided that they can release their energy $2eV$ into the environment, which in our case consists of the transmission line with the side-coupled circuit. Therefore, the dc current reflects both the elastic and inelastic scattering properties of photons. 

Let us consider the current-voltage characteristic in more detail: In the harmonic regime, the effective impedance of the environment is almost flat, except at frequencies near the plasma frequency $\omega_s$ of the superconducting side circuit (the Cooper pair box). 
This results in a feature at $2eV=\hbar\omega_s$ in the current-voltage characteristic. Anharmonic corrections cause additional features near the voltages $2eV=n\hbar\omega_s$ ($n\geq 2$ integer) through inelastic photon scattering. These additional features are most pronounced in the strong coupling regime when $Z$ approaches $R_Q$ ($Z\lesssim R_Q$).

The paper is organized as follows. In Sec.\ II, we introduce the Hamiltonian that describes the circuit studied. The current-voltage characteristic of the probe junction in the harmonic regime is calculated in Sec.\ III.   In Sec.\ IV, we  include a weak anharmonic correction and study the effect of the resulting photon-photon interaction on the current. 
Finally, we conclude in Sec.V.



\section{The circuit studied}
\label{sec-model}

\begin{figure}[t]
\centering
\includegraphics[width=\columnwidth]{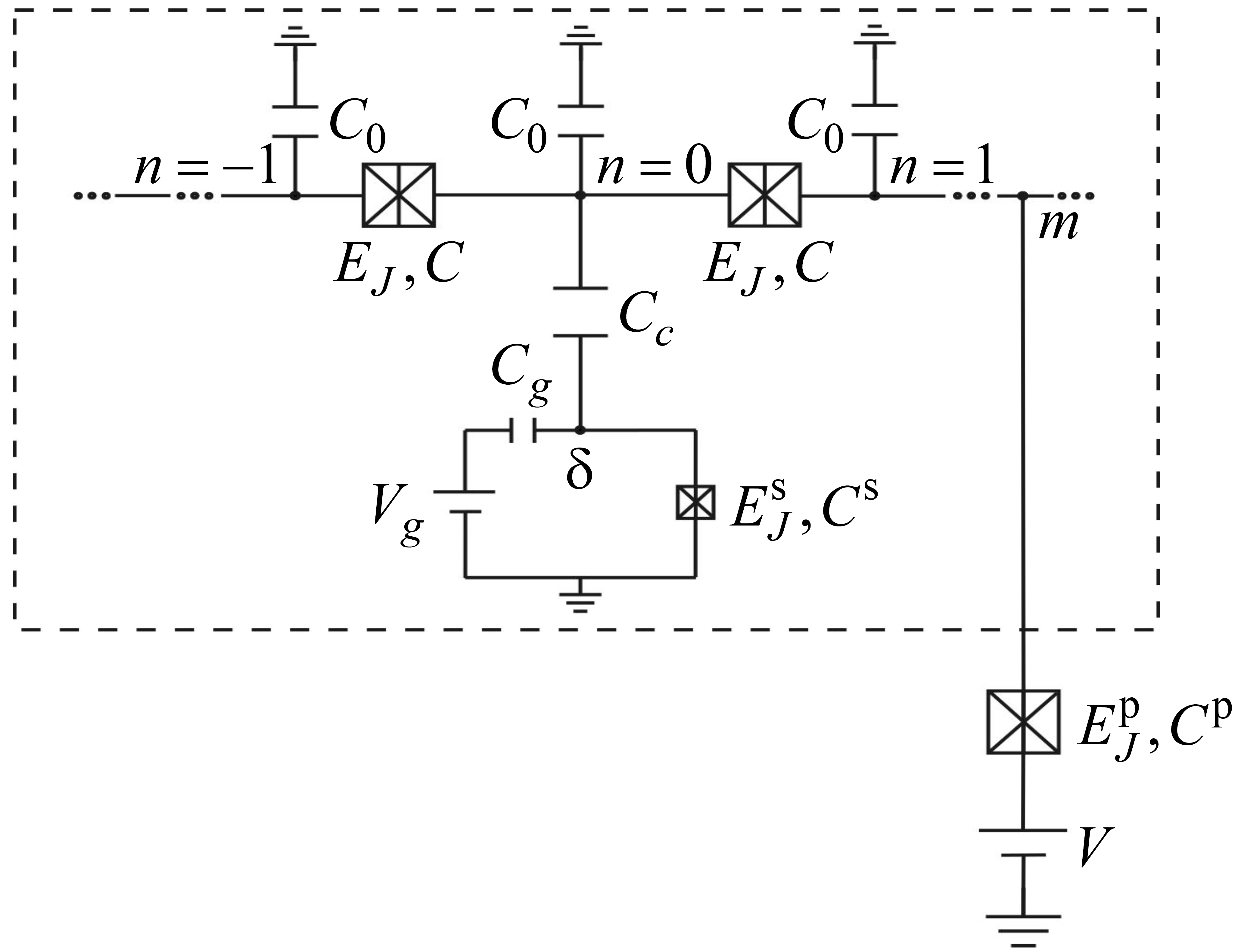}
\caption{The system consists of a transmission line that is capacitively coupled (capacitor $C_c$) to a Josephson junction, shown inside the dashed box. The transmission line is realized using a chain of Josephson junctions with Josephson energy $E_J$ much larger than the charging energy $E_C$.  The system is probed at node $m$ using another Josephson junction (outside the dashed box) whose current-voltage characteristic is sensitive to the properties of the system. \label{fig-setup}}
\end{figure}

We are interested in the interactions of photons propagating in a non-linear electromagnetic environment. In particular, we study a transmission line, consisting of a chain of Josephson junctions, to which an additional Josephson junction acting as the non-linear element is side-coupled at node $n=0$ as shown in Fig.~\ref{fig-setup} (dashed box). We assume weak coupling, namely the coupling capacitance, $C_c$, is much smaller than the characteristic capacitances of the chain and the non-linear element.

The Hamiltonian of the system is, thus, assembled from three parts,
\begin{eqnarray}
H = H_{T} + H_{J} + H_{c}, \label{1}
\end{eqnarray} 
where $H_T$ is the Hamiltonian of the transmission line, $H_J$ is the Hamiltonian of the side-coupled Josephson junction, and $H_{c}$ is the coupling Hamiltonian. 

A transmission line with large impedance can be realized using a chain of Josephson junctions in the limit where the Josephson energy $E_J$ is much larger than the charging energy $E_C$ \cite{MaslukDevoretPRL12,BellGershensonPRL12}. The chain is described by the charge and phase operators at each node $n$, denoted $Q_n$ and $\phi_n$, respectively. They are conjugate variables satisfying the commutation relation $\left[Q_n, \phi_m \right]=-2ie \delta_{nm}$.   
   As $E_J\gg E_C$, phase fluctuations are small and we may approximate the Josephson coupling by a quadratic term.
We further consider the case where the capacitance to the ground $C_0$ is much larger than the mutual capacitance $C$. Then, for frequencies much smaller than the plasma frequency of Josephson junctions in the chain, the Hamiltonian takes the simple form 
\cite{BradleyDoniachPRB84}
\begin{eqnarray}
H_{T}=\sum_n\left[ \frac{Q^2_n}{2C_0} + \frac1{(2e)^2} \frac{(\phi_n - \phi_{n+1})^2}{2L}\right],
\label{2}
\end{eqnarray}
where the  inductance is $L=1/(4e^2E_J)$. Note that we use units where $\hbar=1$. At frequencies $\omega\ll\omega_0$, where $\omega_0 \equiv 1/\sqrt{LC_0}$, the transmission line has a linear spectrum. 

The side-coupled Josephson junction with Josephson energy $E^{\rm s}_J$ is described by the Hamiltonian
\begin{eqnarray}
H_{J}=\frac{(Q_{\delta}+C_g V_g)^2}{2C_{\Sigma}}-E^{\rm s}_J\cos \phi_{\delta}\ ,
\label{3}
\end{eqnarray}
where $Q_{\delta}$ and $\phi_{\delta}$ are the conjugate charge and phase operators at node $\delta$ (see Fig.~\ref{fig-setup}). Furthermore, $C_g$ and $V_g$ are the gate capacitance and gate voltage, respectively, 
{ and $C_{\Sigma}=C^{\rm s}+C_g$ is the total capacitance of the side-coupled Josephson junction.

Finally, we turn to the coupling Hamiltonian $H_c$. When the coupling capacitance is small, $C_c \ll C_0, C_{\Sigma}$, the coupling Hamiltonian reads
\begin{eqnarray}
H_{c}=\frac{C_c}{C_0 C_{\Sigma}}Q_{0}(Q_{\delta}+C_g V_g),
\label{4}
\end{eqnarray}
where we used the fact that for $C\ll C_0$ the coupling is local, i.e., the side-coupled Josephson junction couples only to the charge $Q_0$ at $n=0$.
The Hamiltonian $H$ fully describes our non-linear system. 


As a next step, we introduce the probe circuit used to characterize the photon-photon interactions generated by the non-linear system. The probe circuit consists of yet another Josephson junction, with Josephson energy $E^{\rm p}_J$ and in series with a voltage source as shown in Fig.~\ref{fig-setup}, coupled to the transmission line at node $m$  \cite{Heikkila2004,Basko2013}. The current-voltage characteristic of the probe Josephson junction is influenced by the correlations of the phase $\phi_m(t)$ at node $m$, correlations that depend on the fluctuations in the non-linear environment. The $I$-$V$ characteristic may, thus, be used to characterize the photon-photon interactions in the non-linear system. 

In particular, using $P(E)$-theory, it can be shown that at zero temperature  the current flowing through the probe Josephson junction takes the form \cite{IngoldNazarov92}
\begin{eqnarray}
I(V)=\pi e \left(E^{\rm p}_{J}\right)^2P(2eV),\label{5}
\end{eqnarray}
for voltages $eV \! <2\Delta$, where $\Delta$ is the superconducting gap, and
\begin{eqnarray}
P(E)=\frac{1}{2 \pi } \int\limits^{\infty}_{-\infty} dt \;e^{i Et} \langle e^{i\phi_m(t)}e^{-i\phi_m(0)} \rangle_{H_{\rm env}} \label{6}
\end{eqnarray} 
is the probability of the probe Josephson junction to emit energy $E$ to its environment, described by the Hamiltonian $H_{\rm env}$. Though $P(E)$-theory is usually presented in the context of a linear environment ($H_{\rm env}$ is assumed to be quadratic) \cite{IngoldNazarov92,JezouinPierre13,JoyezPRL13,SouquetPRB13}, Eqs.~\eqref{5}-\eqref{6} hold more generally for a non-linear environment \cite{Heikkila2004}.
If the capacitance of the probe Josephson junction is sufficiently small, $C^{\rm p}\ll C_0$, the Hamiltonian of the environment in Eq.~\eqref{6} may be replaced by the Hamiltonian $H$ of the non-linear system we want to characterize. Our task is then to compute the phase correlator $\langle e^{i\phi_m(t)}e^{-i\phi_m(0)} \rangle_{H}$.

\section{The linear regime}
\label{sec-linear}

As a first step, we will consider the system in the linear regime, where photons do not interact. That is, we assume $E^{\rm
 s}_J\gg e^2/(2C_\Sigma)$ and approximate the junction Hamiltonian $H_J$ in Eq.~\eqref{3} by
\begin{eqnarray}
H^{(0)}_{J}=\frac{(Q_{\delta}+C_g V_g)^2}{2C_{\Sigma}}+\frac{E^{\rm s}_J}{2} \phi^2_{\delta}.
\label{7}
\end{eqnarray}
In this section, we study the behavior of this simplified system described by $H^{(0)}=H_T+H_J^{(0)}+H_c$ to set the basis for investigating interaction effects, the main focus of our work, in the following section.
In this regime, the gate voltage $V_g$ can be gauged out of the Hamiltonian, and the side-coupled circuit behaves as an harmonic oscillator with plasma frequency $\omega_s \equiv 2e \sqrt{E^{\rm s}_J /C_{\Sigma}}$.  
  We will assume that $\omega_s\ll \omega_0$. 

\subsection{Phase-phase correlator}

As the system is non-interacting, the phase-phase correlation function in Eq.~\eqref{6} can be simplified by exploiting Wick's theorem \cite{IngoldNazarov92}:
\begin{eqnarray}
\langle e^{i\phi_m(t)}e^{-i\phi_m(0)} \rangle_{H^{(0)}} = e^{J(t)},  \label{8}
\end{eqnarray}
where
\begin{eqnarray}
 J(t) \equiv \langle  \left[\phi_m(t)-\phi_m(0) \right] \phi_m(0) \rangle_{H^{(0)}}. \label{9}
\end{eqnarray}
To evaluate the correlator, we use the retarded Green function $G^{(0)}_R(\phi_n,\phi_m;t)=i\Theta(t) \langle \left[\phi_n(t), \phi_m(0) \right] \rangle_{H^{(0)}}$, where $\Theta(t)$ is the Heaviside step function. The relation between the two is most easily written in frequency space. At zero temperature, it reads
\begin{eqnarray}
\langle  \phi_n(t) \phi_n(0) \rangle_{H^{(0)}} = 2\int\limits_{0}^\infty\frac{d\omega}{2\pi}\, e^{-i\omega t}\, \Im\! \left[G^{(0)}_R( \phi_n ,\phi_n; \omega)\right].\;\;\;
\label{10}
\end{eqnarray}

The local Green function $G^{(0)}_R( \phi_m, \phi_m;\omega)$ needed to compute $J(t)$ is obtained by deriving its equation of motion and using scattering theory (see Appendix A).
Photons propagate freely in the transmission line and are scattered by the side-coupled 
{harmonic oscillator} 
at node $n=0$, yielding the reflection coefficient
\begin{eqnarray}
r(\omega)=-\left[ 1-2i\frac{\omega_0}{\omega}\left(1+\frac{C_0 C_{\Sigma}}{C^2_c}\frac{\omega^2-{\omega_s}^2}{\omega^2}\right) \right]^{-1} \label{12}.
\end{eqnarray}
In terms of this reflection coefficient, the Green function is 
\begin{eqnarray}
G_R^{(0)} \left( \phi_m, \phi_m;\omega \right)
&=& i\frac{\pi}{\omega} \frac{Z_0}{R_Q} \left[ 1+r(\omega)e^{2i\frac{\omega}{\omega_0}m}\right],\label{11}
\end{eqnarray}
where $Z_0=\sqrt{L/C_0}$ is the impedance of the chain. Under the conditions specified above, $C_0C_\Sigma/C_c^2\gg1$ and $\omega_s\ll\omega_0$, the reflection coefficient has a narrow resonance at $\omega=\omega_s$ with width 
\begin{eqnarray}
\Gamma=\frac{1}{4} \frac{C^2_c}{C_0 C_{\Sigma}} \frac{\omega_s}{\omega_0} \omega_s.
\label{eq:defGamma}
\end{eqnarray}
Close to the resonance, we can approximate Eq.~\eqref{12} as $r(\omega)=-1/[1-i(\omega-\omega_s)/\Gamma]$.

Substituting Eq.~\eqref{10} into Eq.~\eqref{9} and using the Green function \eqref{11}, one obtains
\begin{eqnarray}
J(t) &=&\frac2{R_Q} \int\limits^{\infty}_{0} \frac{d\omega}{\omega} \Re \left[Z(\omega)\right](e^{-i\omega t}-1),  \label{13}
\end{eqnarray}
as expected from $P(E)$-theory \cite{IngoldNazarov92}, with the impedance
\begin{eqnarray}
Z(\omega)=\frac{Z_0}{2} \left[1+r(\omega)e^{2i\frac{\omega}{\omega_0}m}\right].\label{14}
\end{eqnarray}
The prefactor 1/2 corresponds to the fact that the probe junction `sees' an environment consisting of {\it two} half-infinite transmission lines.
Far from the resonance at $\omega_s$, the impedance is unaffected by the side-coupled Josephson junction as $r(\omega)\to0$. In contrast, at the resonance, photons are strongly scattered. In particular when the probe and the scatterer are coupled to the same node ($m=0$), $r(\omega_s)=-1$ so that transport is completely blocked due to destructive interference. Changing the distance between the probe and the scatterer modulates the phase difference between incoming and reflected photons and, thus, creates an interference pattern.

\subsection{Current-voltage characteristic}

To compute the current-voltage characteristic, we need to determine $P(E)$. This can be done numerically using the integral equation~\cite{IngoldNazarov92}
\begin{eqnarray}
E P(E)= 2 \int\limits^{E}_{0} \! dE' \,\frac{\Re \left[Z(E-E') \right]}{R_Q} \, P(E').\label{16}
\end{eqnarray}
The result can be obtained by starting with an arbitrary value $P(0)$ and then using the normalization condition $\int^{E_\textrm{cut-off}}_{0}P(E)dE=1$, where $E_\textrm{cut-off}\gg\omega_s$. 

\begin{figure}[t]
\centering
\includegraphics[width=\columnwidth]{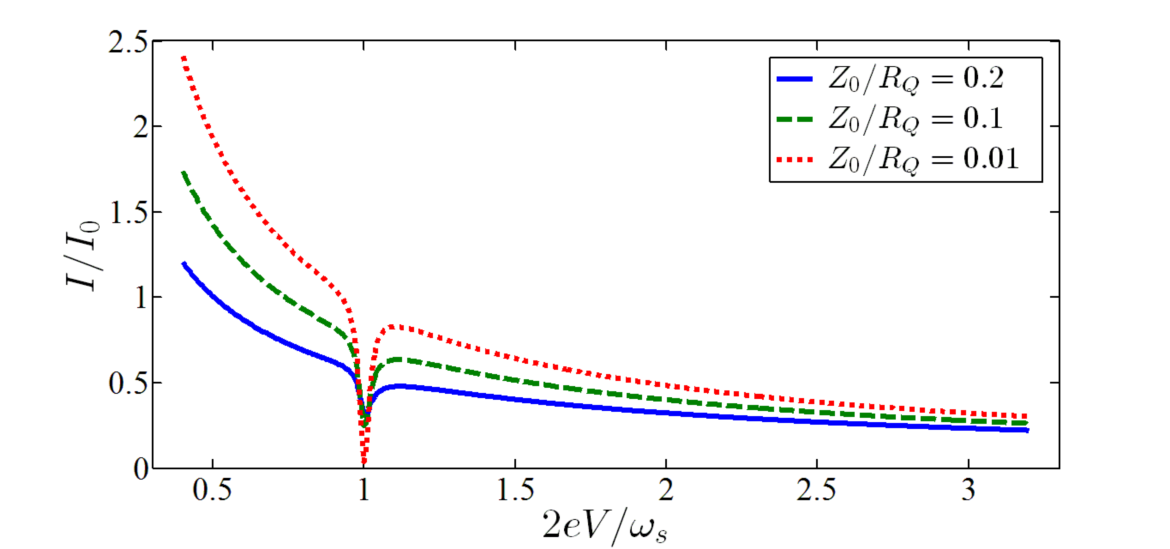}
\caption{(Color online) The linear regime: Current-voltage characteristic of the probe Josephson junction when placed at $m=0$. The parameters are $\Gamma/\omega_s=0.02$, $E_\textrm{cut-off}/\omega_s=20$, and different $Z_0$ ($Z_0/R_Q=0.01$, $0.1$, $0.2$). The side-coupled Josephson junction causes a resonance at $2eV=\omega_s$. In the limit $Z_0/R_Q \to 0$, the current vanishes at the resonance. \label{fig-IV}}
\end{figure}

The current-voltage characteristic is plotted in Fig.~\ref{fig-IV} for several values of the impedance of the transmission line.  The characteristic current is given by 
\begin{eqnarray}
I_0=\frac{\pi e (E^{\rm p}_J)^2}{\omega_s} \frac{Z_0}{R_Q}.
\end{eqnarray}
The background current decreases with increasing voltage. In addition, there is a clear resonance feature at $2eV=\omega_s$.

This result can be understood as follows. The starting point is to recognize that when a bias voltage $V$ is applied, Cooper pairs can flow through the probe junction provided that they can release their energy by emitting one or several photons into the environment.

\begin{figure}[t]
\includegraphics[width=\columnwidth]{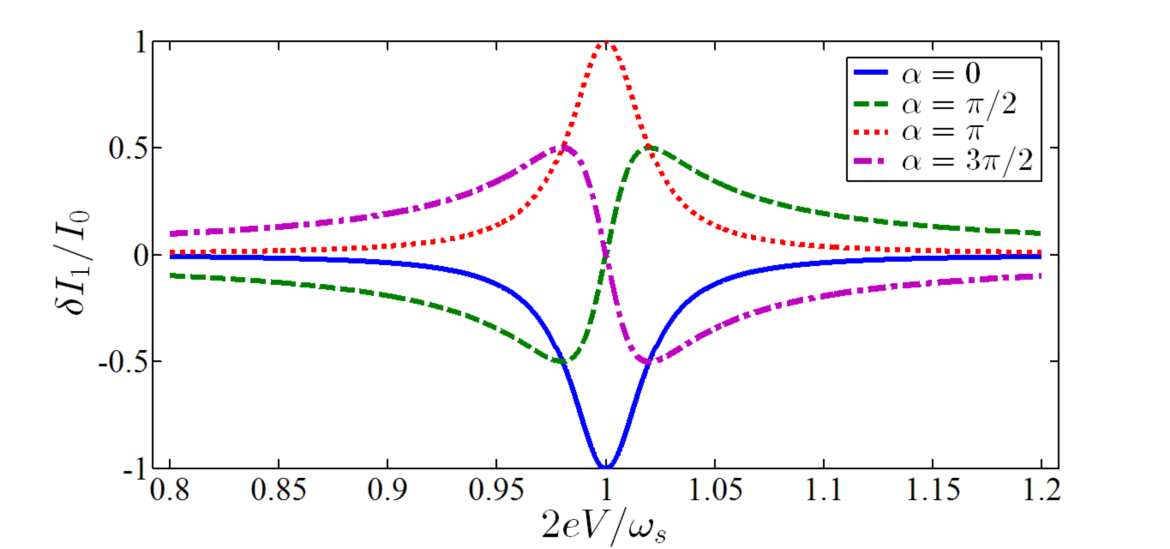}
\caption{(Color online) The resonance in the current-voltage characteristic for different values of the distance between the side-coupled Josephson junction and the probe Josephson junction, $\alpha=2m\omega_s/\omega_0$. Results in the single-photon, linear regime are plotted [Eq.~\eqref{31}] with $\Gamma/\omega_s=0.02$. Note the effect of interference on the shape of the resonance. 
\label{fig-peak1}}
\end{figure}

First, let us concentrate on the regime $Z_0/R_Q\ll1$. In that case, multi-photon processes are suppressed, and we can expand $e^{J(t)}\simeq 1+J(t)$. Thus, the current is proportional to the Fourier transform of $J(t)$ at frequency $2eV$. It is straightforward to show that for a constant impedance this yields a current that decays with increasing voltage as $I(V)\propto 1/V$.  On top of this, the resonance in the impedance at $\omega_s$ due to the side-coupled Josephson junction leads to a resonance in the current-voltage characteristic at $2eV=\omega_s$. Namely, the correction to the current $\delta I_{1}(\delta V)$ at voltages $V=\omega_s/(2e)+\delta V$ takes the form
\begin{eqnarray}
\frac{\delta I_{1}(\delta V)}{I_0}&=& \frac{-\Gamma^2}{(2e\delta V)^2+\Gamma^2} \left(\cos \alpha- \frac{2e\delta V}\Gamma\sin\alpha\right),\label{31}
\end{eqnarray}
where $\alpha=2m\omega_s/\omega_0$.
    This leads to a complete extinction of the current at $\delta V=0$ (at the one photon level) when the probe is coupled to the same node as the side-coupled Josephson junction ($m=0$). The shape of the resonance for different $\alpha$ is shown in Fig.~\ref{fig-peak1}; note the sensitivity to the placement of the probe produced by interference effects. The width of the resonance is given by $W_1=\Gamma/e$ where $\Gamma$ is given in Eq.~\eqref{eq:defGamma}.

Let us now turn to multi-photon processes corresponding to higher order terms in $J(t)$. These processes modify the resonance at $2eV=\omega_s$. In particular, while the scattering from the side-coupled Josephson junction may completely block the single-photon process at that voltage, this is not the case for the multi-photon processes: at most one photon can be on resonance, whereas the other photons will be off resonance and therefore propagate freely. Thus, the multi-photon processes lead to a finite current at the resonance. As an $n$-photon process yields a current contribution proportional to $(Z_0/R_Q)^n$, the resonant structure weakens with increasing $Z_0/R_Q$ due to the increasing importance of multi-photon processes.

In addition, one might expect that multi-photon processes lead to higher order resonances at voltages $2eV=n\omega_s\, (n \geq 2)$. We find, however, that this is not the case. While $2eV=n\omega_s$ is indeed a resonance condition for an $n$-photon process, the non-resonant background from the entire frequency range is large enough to completely overwhelm that contribution. 

Thus, in the linear regime where photons do not interact, the side-coupled Josephson junction leads to a single resonance in the current voltage characteristic at $2eV=\omega_s$. As we will show below, additional features at $2eV=n\omega_s$ with $n \geq 2$ are a signature of photon-photon interactions.

\section{The non-linear regime}

To investigate photon-photon interactions, we now take into account the non-linearity of the side-coupled Josephson junction. In particular, we concentrate on the case of weak non-linearity in the regime $E^{\rm s}_J\gg e^2/(2C_\Sigma)$. To do so, we expand Eq.~\eqref{3} up to fourth order in $\phi_\delta$, 
\begin{eqnarray}
H_J \approx H^{(0)}_J + V,
\label{17}
\end{eqnarray}
where 
\begin{eqnarray}
V=-\frac{E^{\rm s}_J}{24} \phi^4_{\delta}.
\label{18}
\end{eqnarray}
In the following, we treat $V$ as a perturbation.

\subsection{Phase-phase correlator}
 
As the Hamiltonian $H_{\rm nl}\doteq H^{(0)}+V$ describes an interacting system, we can no longer write a closed form expression for the phase-phase correlator \eqref{6} in terms of $\langle  \phi_n(t) \phi_m(0) \rangle$. Instead we expand \eqref{6} in powers of $\phi_m$ as follows,
\begin{widetext}
\begin{eqnarray}\label{19}
\langle e^{i\phi_m(t)}e^{-i\phi_m(0)} \rangle_{H_{\rm nl}} &=& 1 + \langle [\phi_m(t)-\phi_m(0)] \phi_m(0) \rangle_{H_{\rm nl}} + \frac{1}{4}\langle [\phi^2_m(t)-\phi^2_m(0)] \phi^2_m(0) \rangle_{H_{\rm nl}}\\
&& - \frac{1}{6}\left\{\langle  [\phi^3_m(t)-\phi^3_m(0)] \phi_m(0) \rangle_{H_{\rm nl}} +\langle [\phi_m(t)-\phi_m(0)] \phi^3_m(0) \rangle_{H_{\rm nl}}\right\} +{\cal O}\!\left(\phi_m^6\right)\!.\nonumber
\end{eqnarray}
\end{widetext}
Here the two-point phase-phase correlator represents single photon processes, whereas the four-point phase-phase correlators represent two photon processes. As before, we will use Green functions to evaluate the correlators. In addition to the two-point Green function $G_R(\phi_m, \phi_m;\omega)$, we now also need the four-point Green functions $G_R(\phi^2_m, \phi^2_m;\omega)$, $G_R(\phi^3_m, \phi_m;\omega)$, and $G_R(\phi_m ,\phi^3_m;\omega)$. In order to 
facilitate doing
perturbation theory in the interaction $V$, we switch to imaginary-time-ordered or Matsubara Green functions, ${\cal G}$. 

Let us first evaluate the two-point Green function ${\cal G}[\phi_n(\tau)\phi_m(0)]$, corresponding to single photon processes. As shown in Appendix B, using the Dyson equation, we can sum up the perturbation series to all orders in the interactions. The representation in terms of Feynman diagrams is shown in Fig.~\ref{fig-Feynman}.  For the corresponding retarded Green function $G_R(\phi_n ,\phi_m; \omega)$, we finally obtain
\begin{eqnarray}
\label{20}
G_R\left(\phi_n, \phi_m;\omega\right)&=&G^{(0)}_R\left(\phi_n ,\phi_m; \omega\right)\\
&+& \frac{G^{(0)}_R\left(\phi_n ,\phi_{\delta}; \omega\right) \frac{E^{\rm s}_J}{2} \langle \phi^2_{\delta}\rangle_{H^{(0)}} G^{(0)}_R\left(\phi_{\delta}, \phi_m;\omega\right)}{1-\frac{E^{\rm s}_J}{2} \langle \phi^2_{\delta}\rangle_{H^{(0)}} G^{(0)}_R\left(\phi_{\delta} ,\phi_{\delta};\omega\right)},\nonumber
\end{eqnarray} where $G_R^{(0)}$ is the Green function in the absence of interactions.

\begin{figure}[t]
\centering
\includegraphics[width=\columnwidth]{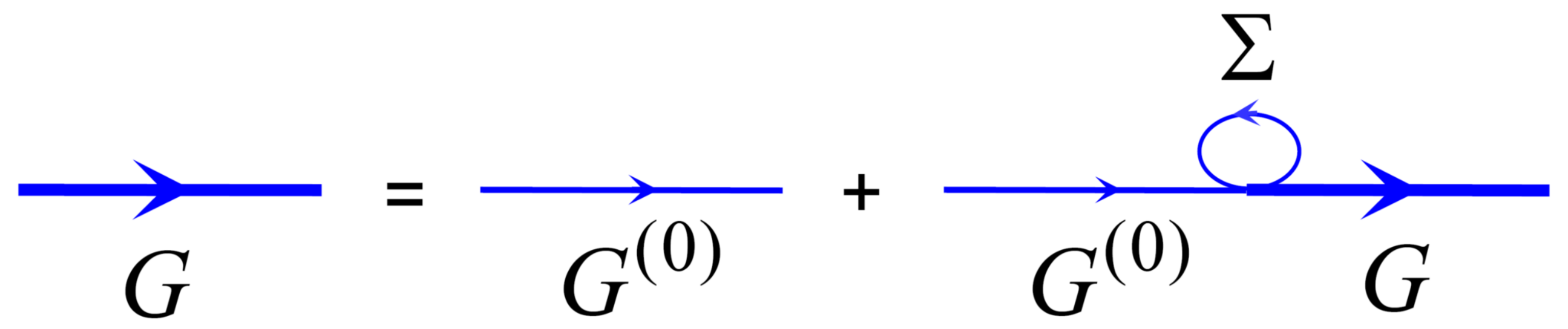}
\caption{Dyson equation for the two-point Green function. The non-linearity results in a self energy, $\Sigma=E^{\rm s}_J\langle \phi^2_{\delta}\rangle_{H^{(0)}} /2$. \label{fig-Feynman}}
\end{figure}

Using the Green functions $G_R^{(0)}$ and $\langle \phi^2_{\delta}\rangle_{H^{(0)}}$ derived in Appendix A, we find 
     that the local Green function preserves its form though with a shifted resonance frequency $\omega_s'$. Namely,
\begin{eqnarray}
G_R \left(\phi_m ,\phi_m;\omega \right) = i \frac{\pi}{\omega} \frac{Z_0}{R_Q} \left[1+r'(\omega) e^{2i\frac{\omega}{\omega_0}m}\right],\label{21}
\end{eqnarray}
where
\begin{eqnarray}
r'(\omega)= -\left[{1-2i\frac{\omega_0}{\omega}\left(1+\frac{C_0 C_{\Sigma}}{C^2_c}\frac{\omega^2-{\omega'_s}^2}{\omega^2}\right)}\right]^{-1}\label{22}
\end{eqnarray}
with $\omega_s'\approx\omega_s[1-\omega_s/(8E^{\rm s}_J)]$.  
In the same way, we can show that this is true for all two-point Green functions. Note that $\delta\omega_s=\omega_s^2/(8E^{\rm s}_J)\ll\omega_s$ coincides with the shift of the excitation energy between the ground and first excited states of the Hamiltonian \eqref{17}.

Next we turn to the four-point Green functions, corresponding to two-photon processes. Using perturbation theory, we may express them in terms of the two-point Green functions. As we saw above, it is essential to sum up the perturbation series to all orders in $V$ to obtain these two-point Green functions. By contrast, we will keep only the lowest order term in $V$ accounting for interactions between the two photons. Then, the four-point Green function   $G_R\left(\phi^2_n, \phi^2_m;\omega\right)$ has two contributions: the first one describes the independent propagation of the two photons, whereas the second one describes the interaction effects. More precisely, the imaginary-time-ordered four-point Green function may be written as ${\cal G}\left[\phi_n^2(\tau) \phi_m^2(0) \right]= {\cal G}^2\left[ \phi_n(\tau) \phi_m(0) \right]+\delta{\cal G}^{\rm int}\left[\phi_n^2(\tau) \phi_m^2(0) \right]$ with
\begin{widetext}
\begin{eqnarray}
\delta{\cal G}^{\rm int}\left[\phi_n^2(\tau) \phi_m^2(0) \right]&\simeq& E^{\rm s}_J \int\limits^{\infty}_{0} d \tau'\; {\cal G}^2\left[ \phi_n(\tau) \phi_{\delta}(\tau') \right] {\cal G}^2\left[ \phi_{\delta}(\tau') \phi_m(0) \right].\label{24}
\end{eqnarray}
The corresponding Feynman diagram is shown in Fig.~\ref{fig-int}(a). 
The expression for the local retarded Green function at zero temperature reads (see Appendix C)\begin{eqnarray}
\delta G_{R}^{\rm int}\left( \phi_m^2, \phi_m^2;\omega \right)&\simeq& \frac{E^{\rm s}_J }{\pi^2} \left(\sum_\pm\int\limits^{\infty}_{0} d\omega_1 \;\Im\left[G_R(\phi_m, \phi_{\delta};\omega_1)\right] G_R(\phi_m, \phi_{\delta};\omega\pm\omega_1)\right)^2.\label{26}
\end{eqnarray} 
The leading order term for the other four-point Green functions $G_{R}\left( \phi_n^3, \phi_m;\omega \right)$ and $G_{R}\left( \phi_n, \phi_m^3;\omega \right)$ is linear in $E_J^{\rm s}$. In particular, we find the local Green functions 
\begin{eqnarray}
G_{R}(\phi^3_m ,\phi_m;\omega)&\simeq& \frac{E^{\rm s}_J }{\pi^2}G_R( \phi_{\delta},\phi_m;\omega)\int\limits^{\infty}_{0} \!d\omega_1 \int\limits^{\infty}_{0} \!d\omega_2\; \Im\left[G_R(\phi_m ,\phi_{\delta};\omega_1)\right] \Im\left[G_R(\phi_m ,\phi_{\delta};\omega_2)\right] \!\!\sum_{s_1,s_2=\pm}\!\!G_R(\phi_m ,\phi_{\delta};\omega+s_1\omega_1+s_2\omega_2)\;\;\;\;\;
\label{27}
\end{eqnarray}
\end{widetext}
and $G_{R}(\phi_m ,\phi^3_m;\omega)=G_{R}(\phi^3_m ,\phi_m;\omega)$. The Feynman diagrams  for the corresponding time-ordered Green functions are shown in Fig.~\ref{fig-int}(b). 

\begin{figure}[b]
\centering
\includegraphics[width=\columnwidth]{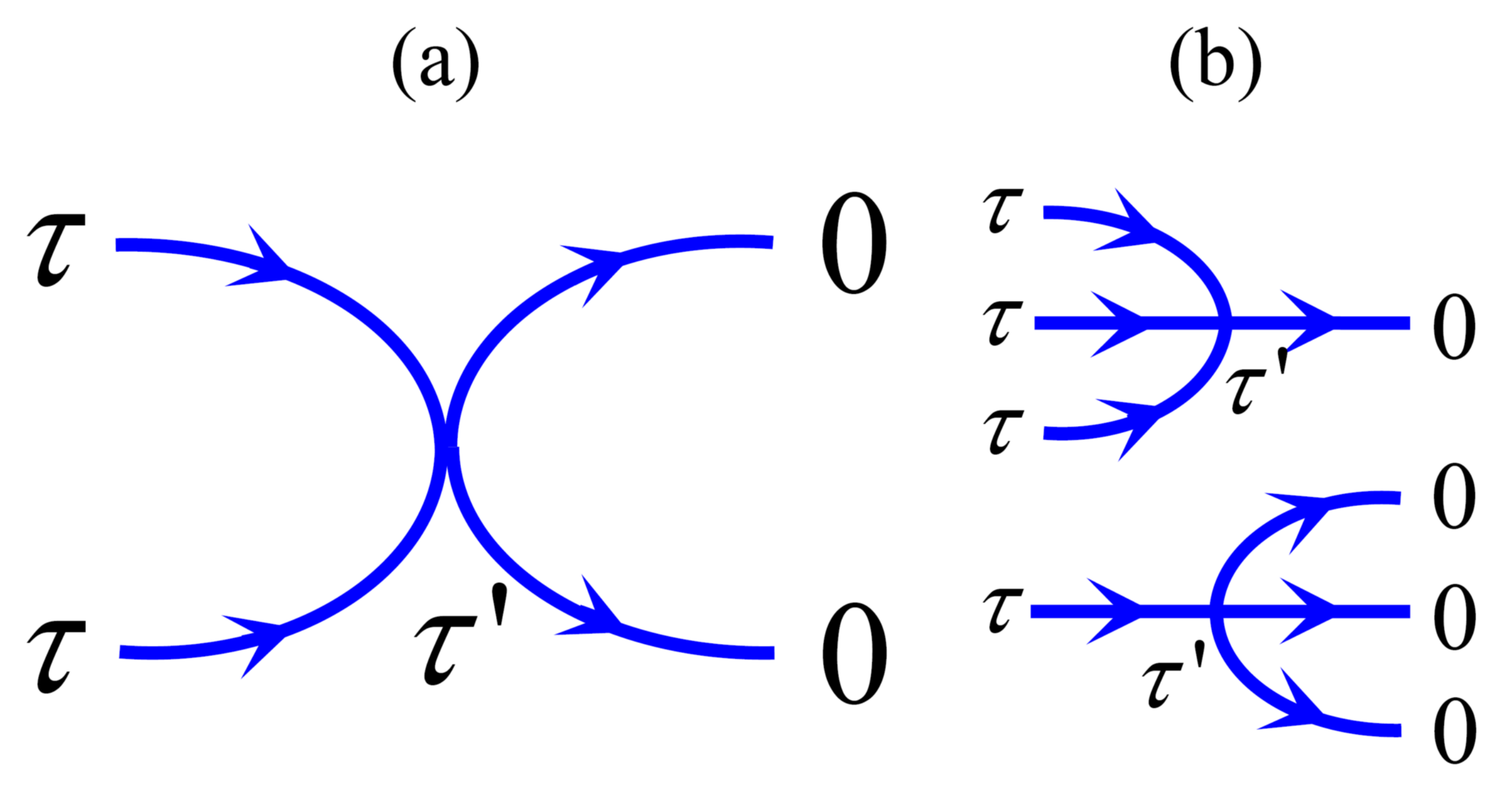}
\caption{The Feynman diagrams for the interaction correction to the four-point Green functions. (a) $\delta {\cal G}^{\rm int}\left[\phi_n^2(\tau) \phi_m^2(0) \right]$. (b)~${\cal G}\left[\phi_n^3(\tau) \phi_m(0) \right]$ and ${\cal G}\left[\phi_n^3(\tau) \phi_m(0) \right]$. \label{fig-int}}
\end{figure}

With the above results we can now write the phase-phase correlator needed to compute $P(E)$ in the following form
\begin{eqnarray}
\langle e^{i\phi_m(t)}e^{\phi_m(0)} \rangle_{H_{\rm nl}} &\simeq& e^{J'(t)}+\delta J^{\rm int}(t)\label{eq-pp-int}
\end{eqnarray}
with 
\begin{eqnarray}
\label{28}
J'(t)&=& \frac{1}{\pi} \int\limits^{\infty}_{0} d \omega\; \Im \left[G_{R}(\phi_m, \phi_m; \omega)\right](e^{-i\omega t}-1)\\
 \delta J^{\rm int}(t)&\simeq&\frac{1}{\pi} \int\limits^{\infty}_{0} d \omega\,\Big\{\frac14 \Im \left[\delta G^{\rm int}_{R}(\phi^2_m ,\phi^2_m; \omega)\right] \\
 &&\qquad-\frac13\Im\left[G_{R}(\phi^3_m, \phi_m;\omega)\right]\Big\}(e^{-i\omega t}-1). \nonumber
\end{eqnarray}

\subsection{Current-voltage characteristic}

Using Eq.~\eqref{eq-pp-int} to compute $P(E)$, we obtain the current
\begin{eqnarray}
I(V)&\simeq& e \left(E^{\rm p}_{J}\right)^2\Big\{\frac12\int\limits^{\infty}_{-\infty} dt \;\exp \left[i 2eVt +J'(t)\right]\\
&+&\frac{1}{4} \Im \left[\delta G^{\rm int}_{R}(\phi^2_m, \phi^2_m; 2eV)\right]-\frac{1}{3}\Im \left[G_{R}(\phi^3_m, \phi_m;2eV)\right]\Big\}.\nonumber
\label{30}
\end{eqnarray}
The first line describes the resonant structure discussed in Sec.~\ref{sec-linear}. Here the only effect of the non-linearity is to shift the resonance from $\omega_s$ to $\omega_s'$. The second line describes interaction effects between two photons. The current-voltage characteristic including these effects is shown in Fig.~\ref{fig-I-int}: it displays additional structure at $2eV=2\omega_s'$.

\begin{figure}[b]
\centering
\includegraphics[width=\columnwidth]{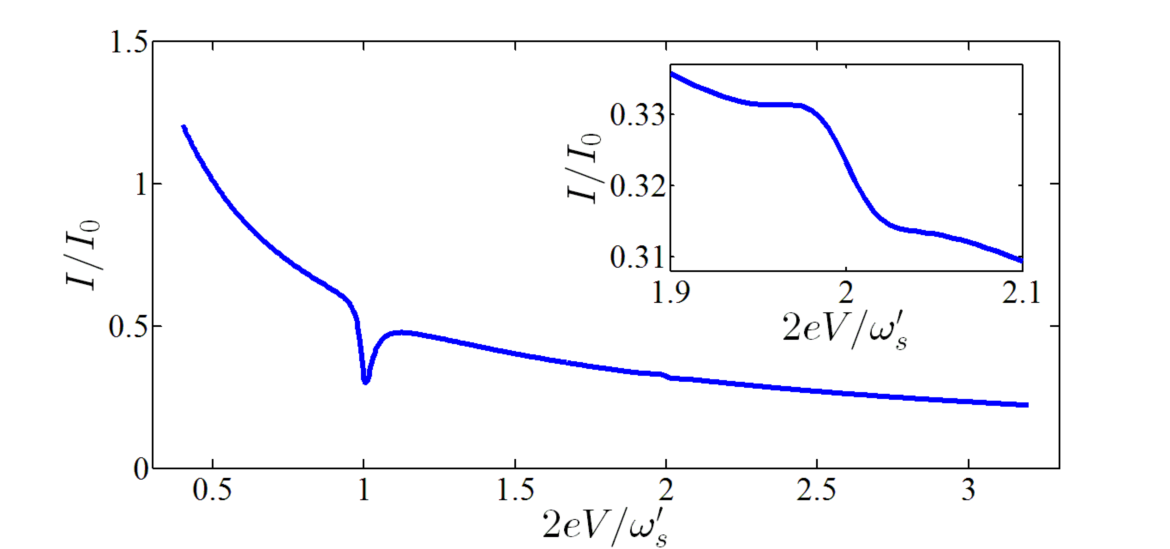}
\caption{The non-linear regime: Current-voltage characteristic of the probe Josephson junction when placed at $m=0$. The parameters are $\omega'_s/E^{\rm s}_J=0.9$, $\Gamma/\omega'_s=0.02$, $E_\textrm{cut-off}/\omega'_s=20$, and $Z_0/R_Q=0.2$. Photon-photon interactions lead to a second resonant feature at $2eV=2\omega'_s$. A zoom on that feature with amplitude $\delta I_2/I_0\propto (Z_0/R_Q)(\omega_s'/E_J^{\rm s})$ is shown in the  inset. \label{fig-I-int}}
\end{figure}

The new peak at $2eV=2\omega_s'$ comes from the contribution $\sim\delta G^{\rm int}_{R}(\phi^2_m, \phi^2_m; 2eV)$. This contribution describes a process in which a Cooper pair tunnels through the probe Josephson junction emitting two photons. When both photons are on resonance with the side-coupled Josephson junction, they interact strongly. This happens when each photon takes away half of the energy of the Cooper pair, $\omega=eV\simeq \omega_s'$. The resulting correction to the current is obtained using Eq.~\eqref{26}. As shown in Appendix C, for voltages $V=\omega_s'/e+\delta V$, it takes the form
\begin{eqnarray}
\label{32}
\delta I_{2}(\delta V)&=& - I_0 \frac{\pi}{32} \frac{Z_0}{R_Q}\frac{\omega_s'}{E_J^{\rm s}}  \frac{\Gamma^2}{[(e\delta V)^2+\Gamma^2]^2} 
\\
&\times & \left\{ \Gamma e\delta V \cos(2\alpha')  - \frac{1}{2}[(e\delta V)^2-\Gamma^2] \sin(2\alpha')\right\},\nonumber
\end{eqnarray}
where $\alpha'=2m\omega_s'/\omega_0$.
 
The characteristic amplitude $A_2$ of the change in current is, thus, much smaller than $I_0$ or the single-photon resonant structure $\delta I_1$, 
\begin{equation}
A_2=\frac{\pi}{64}\frac{Z_0}{R_Q}\frac{\omega'_s}{E^{\rm s}_J} \, I_0\ll I_0.
\label{eq:A2}
\end{equation}
Here, the suppression factor $Z_0/R_Q$ is due to the fact that it is a two-photon process, whereas the suppression factor $\omega_s'/E_J^{\rm s}$ is due to the fact that it is an interaction effect. 
Notice that the widths of the resonances at $2eV= \omega_s'$ and $2eV=2\omega_s'$ are the same. 
The dependence of the shape of the second resonance on the distance $\propto\alpha'$ between the side-coupled Josephson junction and the probe Josephson junction is shown in Fig.~\ref{fig-peak2}. 

\begin{figure}[t]
\centering
\includegraphics[width=\columnwidth]{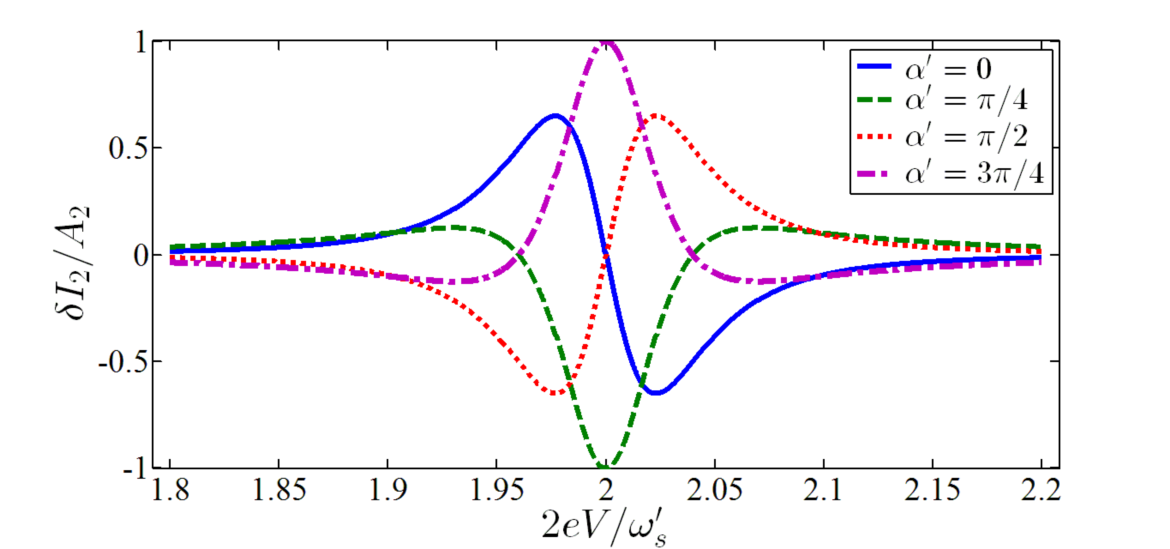}
\caption{The second resonance in the current-voltage characteristic for different values of the distance between the side-coupled Josephson junction and the probe Josephson junction, $\alpha'=2m\omega'_s/\omega_0$. Results are plotted near $2eV=2\omega'_s$ in the two-photon, non-linear regime [Eq.~\eqref{32}] with $\Gamma/\omega'_s=0.02$.
\label{fig-peak2}}
\end{figure}

We finally consider the current contribution stemming from $G_R(\phi^3_m, \phi_m;\omega)$. While it is in principle of the same order as the current contribution from $\delta G^{\rm int}_R(\phi^2_m, \phi^2_m;\omega)$, i.e., it is proportional to $(Z_0/R_Q)(\omega_s'/E_J^{\rm s})I_0$, in this case it is impossible to fulfill the resonance condition simultaneously for all the photons involved. Therefore, this contribution acquires an additional suppression factor $\Gamma/\omega_s'$, and we can neglect it. 

The main interaction effect is, thus, the appearance of a resonance at $2eV=2\omega_s'$ due to two-photon processes. Higher order processes are expected to lead to additional features at $2eV=n\omega_s' \,(n\geq3)$. However, their amplitude decreases rapidly with increasing $n$ and may be estimated as $A_n\sim [(Z_0/R_Q)(\omega_s'/E_J^{\rm s})]^{n-1}I_0 \ll A_2$.

\section{Conclusion}

We have shown that the dc current-voltage characteristic of a Josephson junction provides a sensitive probe to study photon-photon interactions in a non-linear environment. In particular, we investigated the case of a transmission line side-coupled to another Josephson junction whose non-linearity leads to local photon-photon interactions. Scattering of individual photons by the side-coupled Josephson junction results in a resonance feature in the current-voltage characteristic of the probe Josephson junction at $2eV=\omega_s'$, where $\omega_s'$ is the plasma frequency of the side-coupled Josephson junction. By contrast, the interactions due to the non-linearity yield an additional resonance feature at $2eV=2\omega_s'$ due to two-photon processes. Such a feature is thus a clear indication of photon-photon interactions. While we concentrated here on the regime of a weak non-linearity, it will be interesting to see how these features are modified in the strongly non-linear regime. 

\begin{acknowledgments}

The work of Li-Jing Jin is supported by the China Scholarship Council. 
MH and JSM acknowledge support by ANR through grants ANR-11-JS04-003-01 and ANR-12-BS04-0016-03, and an EU-FP7 Marie Curie IRG. HUB acknowledges support from U.S.\ NSF Grant No.~PHY-14-04125. FH is supported by Institut universitaire de France and by the European Research Council (grant no. 306731).
We thank the Fondation Nanosiences de Grenoble for facilitating the exchange between Grenoble and Duke.
\end{acknowledgments}

\appendix

\section{Two point retarded Green functions}

In this appendix, we derive the two-point retarded Green functions $G_R^{(0)}$ of the linear system using equations of motion and scattering theory.

We start with the coupled equations of motion for $G_R^{(0)} \left(\phi_n, \phi_m; \omega \right)$ and $G_R^{(0)} \left(\phi_\delta, \phi_m; \omega \right)$, 
\begin{widetext}
\begin{eqnarray}
\omega^2 G_R^{(0)} \left( \phi_n, \phi_m; \omega \right)-\omega_0^2\left(1+\frac{C_c^2}{C_0C_\Sigma}\delta_{n0}\right) \left[ 2G_R^{(0)} \left( \phi_n, \phi_m; \omega \right)   - G_R^{(0)} \left( \phi_{n+1}, \phi_m; \omega \right)
 -G_R^{(0)} \left( \phi_{n-1}, \phi_m; \omega \right) \right]&&\label{A0}\\- \frac{C_c}{C_0}\omega_s^2 G_R^{(0)} \left( \phi_{\delta}, \phi_m; \omega \right)\delta_{n0}
+ \frac{(2e)^2}{C_0} \delta_{nm}&=&0,\nonumber\\
\left(\omega^2 - \omega_s^2 \right) G_R^{(0)} \left( \phi_{\delta}, \phi_m; \omega \right)- \frac{C_c}{C_\Sigma}  \omega_0^2  \left[ 2G_R^{(0)} \left( \phi_{0}, \phi_m; \omega \right) - G_R^{(0)} \left( \phi_{1}, \phi_m; \omega \right) -G_R^{(0)} \left( \phi_{-1}, \phi_m; \omega \right) \right]&=& 0.\label{A8}
\end{eqnarray}
Combining Eqs.~\eqref{A0} and \eqref{A8} then yields the following equation for $G_R^{(0)} \left( \phi_n, \phi_m; \omega \right)$,
\begin{eqnarray}
\omega^2 G_R^{(0)}\! \left( \phi_n, \phi_m; \omega \right)-\omega_0^2\left(1 \!+\! \frac{C_c^2}{C_0C_\Sigma}  \frac{\omega^2}{\omega^2\!-\!\omega^2_s}
\delta_{n0} \right)
\!\left[ 2G_R^{(0)}\! \left( \phi_n, \phi_m; \omega \right)  \! \!- G_R^{(0)}\! \left( \phi_{n+1}, \phi_m; \omega \right)\! -\!G_R^{(0)}\! \left( \phi_{n-1}, \phi_m; \omega \right) \right]\!
=- \frac{(2e)^2}{C_0} \delta_{nm}.\qquad\label{A1}
\end{eqnarray}
\end{widetext}
If there is no side-coupling, $C_c= 0$, Eq.~\eqref{A1} describes photons propagating freely along the infinite transmission line. At frequencies $\omega\ll\omega_0$, the dispersion is linear, $\omega=\omega_0k$ with wavevector $k\ll1$, and the solution is 
\begin{equation}
G_{R,\,{\rm chain}}^{(0)}\left( \phi_n, \phi_m; \omega \right)=i\frac\pi\omega\frac{Z_0}{R_Q}e^{ik|n-m|}.
\end{equation}

The side-coupling leads to scattering of photons at the node $n=0$. Then, for $m>0$, the solution may be written in the form
\begin{eqnarray}
G_R^{(0)} \left( \phi_n, \phi_m; \omega \right)= \begin{cases}
A e^{ i k n} &    n > m, \\
B \left[e^{ -i k n} +  r(\omega) e^{ i k n}\right] & 0<n<m,\qquad\\
B t(\omega) e^{ -i k n} & n<0,\label{A2}
\end{cases}
\end{eqnarray}
where the reflection and transmission coefficients, $r(\omega)$ and $t(\omega)$, as well as the amplitudes $A$ and $B$ have to be determined using the boundary conditions at $n=0$ and $n=m$. One finds $t(\omega)=1+r(\omega)$ with  
  $r(\omega)$ given by Eq.~\eqref{12} in the main text.  
Furthermore,
\begin{eqnarray}
B&=&i\frac{\pi}{\omega} \frac{Z_0}{R_Q} e^{i km},\label{A5}\\
A&=&B\left[e^{-2ikm}+r(\omega)\right].\label{A6}
\end{eqnarray}
The result is obtained by substituting Eqs.~(\ref{12}), (\ref{A5}), and (\ref{A6}) into Eq.~\eqref{A2}. Generalizing to arbitrary $m$, we find
\begin{eqnarray}
G_R^{(0)} \left( \phi_n, \phi_m; \omega \right)
&=& i\frac{\pi}{\omega} \frac{Z_0}{R_Q} \left[e^{ik|n-m|}+r(\omega)e^{ik(|n|+|m|)}\right].\qquad\label{A7}
\end{eqnarray}
The local Green function needed to evaluate $P(E)$, thus, reads
\begin{eqnarray}
G_R^{(0)} \left( \phi_m, \phi_m; \omega \right)
&=& i\frac{\pi}{\omega} \frac{Z_0}{R_Q} \left[1+r(\omega)e^{2ik|m|}\right].\quad
\end{eqnarray}

While this is the only Green function needed in the linear case, more Green functions are required in the non-linear case.
Using Eq.~\eqref{A8}, we obtain 
\begin{eqnarray}
G_R^{(0)} \left( \phi_{\delta}, \phi_m; \omega \right) &=& - 2 \frac\pi\omega \frac{1}{R_QC_c\omega}  r(\omega) e^{ik|m|}.\label{A9}
\end{eqnarray}
Similarily, the Green functions $G_R^{(0)} \left( \phi_m, \phi_\delta; \omega \right)$ and  $G_R^{(0)} \left( \phi_{\delta}, \phi_{\delta}; \omega \right)$ obey coupled equations of motion. One may show that $G_R^{(0)} \left( \phi_m, \phi_\delta; \omega \right)=G_R^{(0)} \left( \phi_{\delta}, \phi_m; \omega \right)$, whereas the equation for $G_R^{(0)} \left( \phi_{\delta}, \phi_{\delta}; \omega \right)$ is 
\begin{widetext}
\begin{eqnarray}
\left(\omega^2 - \omega_s^2 \right) G_R^{(0)} \left( \phi_{\delta}, \phi_{\delta}; \omega \right)- \frac{C_c}{C_\Sigma}  \omega^2_{0}  \left[ 2G_R^{(0)} \left( \phi_{0}, \phi_{\delta}; \omega \right) - G_R^{(0)} \left( \phi_{1}, \phi_{\delta}; \omega \right) - G_R^{(0)} \left( \phi_{-1}, \phi_{\delta}; \omega \right) \right]= -  \frac{(2e)^2}{C_{\Sigma}}. \label{A14}
\end{eqnarray}
\end{widetext}
Using Eq.~\eqref{A9}, one obtains
\begin{eqnarray}
G_R^{(0)} \left( \phi_{\delta}, \phi_{\delta}; \omega \right) &=& - 4i \frac\pi\omega\frac1{R_QZ_0(C_c\omega)^2}r(\omega).\label{A15}
\end{eqnarray}
Using the explicit expresssion for $r(\omega)$, Eq.~\eqref{A15} may be rewritten as
\begin{eqnarray}
G_R^{(0)} \left( \phi_{\delta}, \phi_{\delta}; \omega \right) &=&- \frac{2\pi}{R_QC_\Sigma}\frac1{\omega^2 -\omega^2_s+i\frac{C_c^2}{2C_0C_\Sigma}\frac{\omega^2}{\omega_0}(\omega-2i\omega_0)}.
\nonumber\\
\end{eqnarray}
Finally, using the fact that $C_c^2/(C_0C_\Sigma)\ll1$, we approximate
\begin{eqnarray}
G_R^{(0)} \left( \phi_{\delta}, \phi_{\delta}; \omega \right) &\simeq&- \frac{2\pi}{R_QC_\Sigma}\frac{1}{\omega^2 - (\omega_{s} - i \Gamma)^2 }, \label{A17}
\end{eqnarray}
where 
$$\Gamma =  \frac{1}{4} \frac{C_c^2}{C_0 C_{\Sigma}} \frac{{\omega_s}^2}{\omega_0}.$$
This result also allows us to evaluate
\begin{eqnarray}
 \langle \phi_{\delta}^2 \rangle_{H^{(0)}} = \frac{1}{\pi} \int\limits ^{\infty}_{0} d\omega\; \Im \left[ G^{(0)}_{R}\left( \phi_{\delta}, \phi_{\delta}; \omega \right) \right] 
 = \frac{\omega_s}{2E^{\rm s}_J}.\quad\label{A19}
\end{eqnarray}

\section{The Dyson equation}

In the following, we present the derivation of Eq.~\eqref{20}. 

The time-ordered two-point Green function is defined as
\begin{eqnarray}
{\cal G}_{\tau} \left[ \phi_n(\tau)\phi_m(0)\right]= \langle T_{\tau} \phi_n(\tau)\phi_m(0) \rangle_{H_{\rm nl}},\label{B1}
\end{eqnarray}
where $ T_{\tau}$ is the time-ordering operator. Eq.~\eqref{B1} can be rewritten as
\begin{eqnarray}
{\cal G}_{\tau} \left[ \phi_n(\tau)\phi_m(0)\right]=\frac{\langle T_{\tau} \phi_n(\tau)\phi_m(0)S(\infty) \rangle_{H^{(0)}}}{\langle S(\infty)\rangle_{H^{(0)}}}, \label{B2}
\end{eqnarray}
where $S(\infty)= T_{\tau} \exp \left[-\int^{\infty}_{0}d\tau'\; V(\tau')\right]$.

Expanding Eq.~\eqref{B2} up to first order in the perturbation $V=-E_J^{\rm s}\phi_\delta^4/24$ and using Wick's theorem yields
\begin{widetext}
\begin{eqnarray}
&&{\cal G}_{\tau} \left[ \phi_n(\tau)\phi_m(0)\right]\simeq  {\cal G}^{(0)}_{\tau}\left[ \phi_n(\tau)\phi_m(0) \right]
+ \int\limits^{\infty}_{0} d\tau' \;{\cal G}^{(0)}_{\tau}\left[ \phi_n(\tau)\phi_{\delta}(\tau') \right] \frac{E^s_J}{2} \langle \phi^2_{\delta} \rangle_{H^{(0)}} {\cal G}^{(0)}_{\tau}\left[\phi_{\delta}(\tau')\phi_m(0) \right].\label{B4}
\end{eqnarray}
\end{widetext}
After Fourier transformation and analytical continuation, one obtains the corresponding retarded Green function,
\begin{eqnarray}
G_{R} \left( \phi_n,\phi_m;\omega \right)&\simeq& G_{R}^{(0)}\!\left( \phi_n,\phi_m;\omega \right)\label{B6}\\
&&+ G_{R}^{(0)}\!\left(\phi_n,\phi_{\delta};\omega\right) \frac{E^{\rm s}_J}{2} \langle \phi^2_{\delta} \rangle_{H^{(0)}} G_{R}^{(0)}\!\left(\phi_{\delta},\phi_m;\omega \right)\!.\nonumber\end{eqnarray}
While far from the resonance at $\omega=\omega_s$ the second term is much smaller than the first one, this is no longer true close to the resonance. Thus, this first order expansion is not sufficient is to describe the modifications to the resonance due to the perturbation.  It is possible to go beyond the first order expansion by realizing that $E_J^{\rm s}\langle \phi^2_{\delta} \rangle_{H^{(0)}} /2$ is a local self-energy, $\Sigma(\phi_\delta,\phi_\delta)$. Thus, one obtains the Dyson equation
\begin{eqnarray}
G_{R} \left(\phi_n,\phi_m;\omega \right)&=&G_{R}^{(0)}\!\left( \phi_n,\phi_m;\omega \right)\label{B8}\\
&&+ G_{R}^{(0)}\!\left(\phi_n,\phi_{\delta};\omega\right)\frac{E^s_J}{2} \langle \phi^2_{\delta} \rangle_{H^{(0)}} G_{R}\left(\phi_{\delta},\phi_m;\omega \right),\nonumber
\end{eqnarray}
A similar equation can be written for the Green function $G_{R} \left(\phi_{\delta},\phi_m;\omega \right)$. Namely,
\begin{eqnarray}
G_{R} \left(\phi_{\delta},\phi_m;\omega \right)&=&G_{R}^{(0)}\left( \phi_{\delta},\phi_m;\omega \right)\label{B9}\\
&&+ G_{R}^{(0)}\left(\phi_{\delta},\phi_{\delta};\omega\right)\frac{E^s_J}{2} \langle \phi^2_{\delta} \rangle_{H^{(0)}} G_{R}\left(\phi_{\delta},\phi_m;\omega \right). \nonumber
\end{eqnarray}
Combining Eqs.~\eqref{B8} and \eqref{B9}, we obtain the result
\begin{widetext}
\begin{eqnarray}
G_{R} \left( \phi_n,\phi_m;\omega \right)=G_{R}^{(0)}\left( \phi_n,\phi_m;\omega \right)+ \frac{G_{R}^{(0)}\left(\phi_n,\phi_{\delta};\omega\right) \frac{E^s_J}{2} \langle \phi^2_{\delta} \rangle_{H^{(0)}}G_{R}^{(0)}\left(\phi_{\delta},\phi_m;\omega \right)}{1- \frac{E^s_J}{2} \langle \phi^2_{\delta} \rangle_{H^{(0)}}G_{R}^{(0)}\left( \phi_{\delta},\phi_{\delta};\omega\right)}.
 \label{B11}
\end{eqnarray}
\end{widetext}
Then, using the Green functions $G_R^{(0)}$ of the linear problem derived in appendix A, we find that the full Green function $G_{R} \left( \phi_n,\phi_m;\omega \right)$ has the same form as $G_{R}^{(0)} \left( \phi_n,\phi_m;\omega \right)$ though with a shift of the resonance frequency, $\omega_s\to\omega_s'\approx\omega_s\left(1-\langle \phi^2_{\delta} \rangle_{H^{(0)}}/4\right)$. Similarily, we find that this frequency shift appears in all two-point Green functions.

\section{The contribution of photon-photon interaction}


The four-point retarded Green function $\delta G_{R}^{\rm int}\left( \phi_m^2, \phi_m^2;\omega \right)$, needed to compute the interaction contribution to the current-voltage characteristic, is obtained from Eq.~\eqref{24} by taking the Fourier transform and then performing the analytical continuation from Matsubara to real frequencies, $i\omega_{\nu} \to \omega+i0^{+}$, and using standard methods of contour integration. One obtains Eq.~\eqref{26} which takes the form $\delta G_{R}^{\rm int}\left( \phi_m^2, \phi_m^2;\omega \right)\simeq(E^{\rm s}_J/\pi^2)f^2(\omega)$, where 

\begin{eqnarray}
f(\omega)=\sum_\pm\int\limits^{\infty}_{0}\!\! d\omega_1 \,\Im\left[G_R(\phi_m, \phi_{\delta};\omega_1)\right]G_R(\phi_m, \phi_{\delta};\omega\!\pm\!\omega_1).
\nonumber\\
\end{eqnarray} 
The integral is dominated by frequencies where both Green functions are close to resonance, $\omega_1\approx\omega\pm\omega_1\approx\omega_s'$. This requires $\omega\approx 2\omega_s'$. We, thus, approximate $\omega=2\omega_s'+\delta\omega$ and $\omega_1=\omega_s'+\delta\omega_1$. The Green functions close to resonance take the form
\begin{eqnarray}
G_R^{(0)} \left( \phi_{\delta}, \phi_m; \omega_s'+\delta\omega \right) &\simeq& 2 \frac\pi{\omega_s'} \frac{1}{R_QC_c\omega_s'}  \frac1{1-i\frac{\delta\omega}\Gamma} e^{i\frac{\omega_s'}{\omega_0}|m|}.\qquad
\end{eqnarray}
We then rewrite
\begin{widetext}
\begin{eqnarray}
f(\omega)&\simeq&\left(\frac{2\pi}{R_QC_c(\omega_s')^2}\right)^2e^{i\frac{\omega_s'}{\omega_0}|m|}\int\limits_{-\infty}^\infty d\delta\omega_1\;\frac{\sin\frac{\omega_s'|m|}{\omega_0}+\frac{\delta\omega_1}\Gamma\cos\frac{\omega_s'|m|}{\omega_0}}{1+\left(\frac{\delta\omega_1}\Gamma\right)^2}\frac1{1-i\frac{\delta\omega-\delta\omega_1}\Gamma}.
\end{eqnarray}
It is straightforward to evaluate the convolution integrals to obtain
\begin{eqnarray}
\Re\left[f(\omega)\right]&\simeq&\pi\left(\frac{\pi C_c}{2R_QC_0C_\Sigma}\right)^{\!\!2}\,\frac{\delta\omega\cos\alpha'+ 2\Gamma\sin\alpha'}{(\delta \omega)^2+4\Gamma^2},\\
\Im\left[f(\omega)\right]&\simeq&- \pi\left(\frac{\pi C_c}{2R_QC_0C_\Sigma}\right)^{\!\!2}\,\frac{2\Gamma\cos\alpha'- \delta\omega\sin\alpha'}{(\delta \omega)^2+4\Gamma^2}.\quad
\end{eqnarray}
Finally, to compute the current-voltage characteristic, we need
\begin{eqnarray}
\Im\left[\delta G_{R}^{\rm int}\left( \phi_m^2, \phi_m^2;2\omega_s'+\delta\omega\right)\right]&\simeq&\frac{2E^{\rm s}_J}{\pi^2}\Re\left[f(2\omega_s'+\delta\omega)\right]\Im\left[f(2\omega_s'+\delta\omega)\right]\\
&\simeq&-\frac{\pi^2}8\frac1{E_J^{\rm s}}\left(\frac{Z_0}{R_Q}\right)^2\frac{4\Gamma^2}{[(\delta \omega)^2+4\Gamma^2]^2}\left\{2\delta\omega\Gamma\cos(2\alpha')-\frac12\left(\delta\omega^2-4\Gamma^2\right)\sin(2\alpha')\right\}.\nonumber
\end{eqnarray}
\end{widetext}

\bibliography{WQED_2015-05,Dissipation_2015-05}

\end{document}